\documentclass{sig-alternate-05-2015}
\usepackage{color}
\usepackage[breaklinks]{hyperref}
\hypersetup{
      colorlinks=true,
      citecolor=blue,
      linkcolor=black,
      urlcolor=blue
}
\usepackage{times}
\usepackage{latexsym}
\usepackage{amsmath}
\usepackage{url}
\usepackage[mathscr]{euscript}
\usepackage{multirow}
\usepackage{arydshln}
\usepackage{color}
\usepackage{graphicx}
\usepackage{caption}
\usepackage{subcaption}
\usepackage[dvipsnames]{xcolor}
\usepackage{amssymb}

\newcommand{\jdk}[1]{{\bf }\textcolor{black}{#1}{\bf}}

\newcommand{\specialcell}[2][c]{\begin{tabular}[#1]{@{}c@{}}#2\end{tabular}}

\title{Low-dimensional Query Projection based on Divergence Minimization Feedback Model for Ad-hoc Retrieval}

 \author{Javid Dadashkarimi \and Masoud Jalili Sabet \and Heshaam Faili \and Azadeh Shakery \\
         School of ECE, College of Engineering, University of Tehran, Iran \\
       \{dadashkarimi,jalili.masoud,hfaili, shakery\}@ut.ac.ir}

\date{}

\begin{document}

\maketitle

\begin{abstract}
Low-dimensional word vectors have long been used in a wide range of applications in natural language processing.
In this paper we shed light on estimating query vectors in ad-hoc retrieval where a limited information is available in the original query.
Pseudo-relevance feedback (PRF) is a well-known technique for updating query language models and expanding the queries with a number of relevant terms. 
We formulate the query updating in low-dimensional spaces first with rotating the query vector and then with scaling. 
These consequential steps are embedded in a query-specific projection matrix capturing both angle and scaling.
In this paper we propose a new but not the most effective technique necessarily for PRF in language modeling, based on the query projection algorithm.
We learn an embedded coefficient matrix for each query, whose aim is to improve the vector representation of the query by transforming it to a more reliable space, and then update the query language model.
The proposed embedded coefficient divergence minimization model (ECDMM)
takes top-ranked documents retrieved by the query and obtains a couple of positive and negative sample sets;
these samples are used for learning the coefficient matrix which will be used for projecting the query vector and updating the query language model using a softmax function.
Experimental results on several TREC and CLEF data sets in several languages demonstrate effectiveness of ECDMM.
The experimental results reveal that the new formulation for the query works as well as state-of-the-art PRF techniques and outperforms state-of-the-art PRF techniques in a TREC collection in terms of MAP,P@5, and P@10 significantly.
\end{abstract}
\section{Introduction}
Low-dimensional word vectors have long been tailored as useful resources for understanding machine-readable texts.
latent semantic analysis (LSI), probabilistic latent semantic analysis (PLSI), non-negative matrix factorization (NMF), and latent Dirichlet allocation (LDA) are statistical techniques to this end~\cite{Deerwester:1990,Hofmann:1999,Zamani:2016,Blei:2003}.

LSI reduces high-dimensional count vectors to low-dimensional latent semantic space \cite{Deerwester:1990};
PLSI provides a statistical formulation on LSA \cite{Hofmann:1999}; 
NMF is tailored over a term-document weighting matrix for term recommendation and query re-weighting \cite{Zamani:2016};
and LDA assumes Dirichlet allocation for both document-topic and term-topic distributions and introduce a better generative model to this end \cite{Blei:2003}.
Although their effectiveness in low-dimensional text modeling, all the methods rely on the simplified bog of word assumption;
Recently, low-dimensional word vectors constructed from a feed-forward neural networks with single hidden layer have been used successfully for text modeling in a wide range of applications~\cite{Mikolov:2013,Collobert:2011,Hinton:2006}.
The neural network-based vectors captures positions of the words in a context and indirectly embed multiple n-grams in the final model \cite{Mikolov:2013}. 
The constructed vectors have been used in document classification \cite{Kusner:2015}, term-by-term expansion \cite{ALMasri:2016}, sentiment analysis \cite{Socher:2013}, named entity recognition (NER) \cite{Turian:2010}, semantic role labeling \cite{Collobert:2011}, and query re-weighting \cite{Zheng:2015}.
A linear neural network learns a number of hidden variables for a word, capturing both semantic and syntactic information~\cite{Mikolov:2013,Zheng:2015}. 
A further convolutional layer is used for estimating vectors for pieces of texts, sentences, documents, and queries~\cite{Collobert:2011}.
Nevertheless, simple \textit{max}, \textit{min}, and \textit{avg} aggregation layers are used instead successfully \cite{Vulic:2015b,Le:2014}.

Word embedding has limited applications in ad-hoc retrieval but, it is likely to have a burgeon in this area during the next years \cite{Manning:16}.
However, it has been shown that query-document vector similarity, in which there is not document length normalization nor term frequency consideration, degrades the retrieval performance and thus it is recommend to use these vectors within state-of-the-art retrieval frameworks~\cite{Zheng:2015,Vulic:2015b,Dadashkarimi:2016}.
Estimating robust language models for documents and queries is a required task in this area. 
Although it has been proposed a number of interesting works in low-dimensional text modeling  \cite{Ganguly:2015,V:2014}, but 
as far as we know, there is no in-detail study investigating on low-dimensional query language modeling where a few number of keywords being posed by the users.

In this paper we shed light on estimating query vectors in ad-hoc retrieval where a limited information about a user intention is available in the original query.
Therefore it is required to propose a novel technique for building/updating the low-dimensional query vectors which is the focus of this paper.
We formulate the query modeling/updating in low-dimensional spaces first with rotating the query vector and then with scaling. 
These consequential steps can be embedded in a query-specific projection matrix capturing both angle and scaling.
Thus we propose to learn this matrix for each query and then project the query vector to a more relevant low-dimensional space.
Learning this matrix requires low-dimensional vectors of a number of relevant/irrelevant vectors to the query.
To this aim, first we extract a couple of word sets from top-ranked documents retrieved in response to the original query and then find a matrix projecting the query vector close to the relevant vectors and far away from the irrelevant ones.
The projected query vector is then be used for building a query language model after applying a softmax/sigmoid function.
The obtained query language model can be tailored in a statistical language modeling framework, the state-of-the-art retrieval framework.


Experimental results on \jdk{several} data sets from TREC and CLEF in English, French, Spanish, German, and Persian demonstrate the effectiveness of the proposed method. 
The experimental results reveal that the new formulation for the query vector modeling/updating works as well as state-of-the-art PRF techniques and even better in a few number of collections.
The proposed method outperforms all competitive baselines in terms of MAP in a TREC collection and a German collection.
The performances in the French collection and the Spanish one are very competitive to NMF.

\section{Previous Works}
\subsection{Pseudo-relevance Feedback}
\label{Sec:Pseudo-relevance Feedback}
Top-ranked documents $F=\{d_1,d_2,..,d_{|F|}\}$ in response to a query $q=\{q_1,q_2,..,q_m\}$ have long been used as helpful resources for expanding the query~\cite{Lavrenko:01,Lv:2014}.
Lavrenko et al., introduced relevance models for updating the query.~The RM1 method models the query as:

\begin{equation}
 p(w|\theta_q)\!\!\propto \!\! \sum_{d \in F} p(w|\theta_d)p(\theta_d)\prod_{i=1}^{|q|} p(q_i|\theta_d)
 \label{eq:rm1}
\end{equation}

 where $\theta_d$ is the language model of document $d\in F$.
The RM2 models the query in another way as:

\begin{equation}
p(w|\theta_q) \!\!\propto \!\!p(w) \prod_{i=1}^{|q|} \sum_{d \in F} p(q_i|\theta_d) \frac{p(w|\theta_d)p(\theta_d)}{p(w)}.    
\end{equation}

The obtained relevance models can be interpolated with the original query as follows:
\begin{equation}
p(w|\theta'_q) = (1-\alpha)p(w|\theta_q) + \alpha p_{\text{ml}}(w|q)
\label{Eq:linear}
\end{equation}

where $p_{\text{ml}}(w|q)$ is the maximum likelihood estimation of the original query. The interpolated model for RM1 and RM2 are known as RM3 and RM4 respectively~\cite{Jaleel:2004}.

Zhai \& Lafferty in \cite{Zhai:2001} introduced the mixture model (MIXTURE) based on an expectation maximization algorithm for modeling the feedback documents.

 \begin{align}
t^{(n)}(w)\!\! =\!\!\frac{(1-\lambda)p^{(n)}_{\lambda}(w|\theta_F)}{(1-\lambda)p^{(n)}_{\lambda}(w|\theta_F)+\lambda p(w|\mathcal{C})}) \\ 
p^{(n+1)}_{\lambda}(w|\theta_F)\!\!=\!\! \frac{\sum_{j=1}^{n}c(w;d_j)t^{(n)}(w)}{\sum_{i}\sum_{j=1}^{n}c(w_i;d_j)t^{(n)}(w)}
\end{align}

where $t^{(n)}(w)$ is topicality of the word $w$ at $n-$th iteration. $c(w_i;d_j)$ indicates the count of the word $w_i$ in document $d_j$. 
The authors introduced another model for feedback modeling based on divergence minimization in \cite{Zhai:2001} as follows:
\begin{align}
p(w|\hat{\theta}_F) \propto \exp \Big(&\frac{1}{1-\lambda}\frac{1}{|F|}\sum_i\log p(w|\hat{\theta}_{d_i})\nonumber\\
&-\frac{\lambda}{1-\lambda} \log p(w|\mathcal{C})\Big)
\end{align}

where $\lambda$ is a controlling constant. Later Lv \& Zhai modified this framework in  \cite{Lv:2014} and introduced maximum entropy divergence minimization model (MEDMM) that aims to find $\hat{\theta_F}$ as follow:
\begin{equation}
\underset{\theta} {\mathrm{arg~min}}\!
\sum_{d\in F}\! \alpha_d H(\theta_F,\theta_d)\! -\! \lambda H(\theta_F,\theta_C)\! -\! \beta H(\theta_F)
\end{equation}
where $H(\theta_1,\theta_2)$ is the cross-entropy between $\theta_1$ and $\theta_2$ and $H(\theta)$ is the entropy of $\theta$.

Zamani et al., \cite{Zamani:2016} proposed a new technique for PRF based on non-negative matrix factorization. The proposed relevance feedback based on matrix factorization (RFMF) first builds $A \in \mathbb{R}^{m\times n}_+$ based on occurrences of terms in $d\in F$ where $m=|F|+1$ and $n$ is the number of unique words in $F$ (the $|F|+1$-th row belongs to the original query).
Second, RFMF aims to decompose $A^{m\times n}$ to $U \in \mathbb{R}^{m\times r}_+$ and $V \in \mathbb{R}^{r\times n}_+$. 
The final product of $U^{m\times r}$ and $V^{r\times n}$ re-weights the query language model.
\subsection{Low-dimensional Vectors}
Low-dimensional representations of words are tailored in a variety of tasks in natural language processing~\cite{Collobert:2011}. To learn these vectors a common approach is to predict the context of each word and then aim at minimizing a loss function. 
This method is known as skip-gram negative sampling and can be interpreted as a binary regression task as follows:
\begin{equation}
\underset{\theta} {\mathrm{arg~min}}\! \!\!\!\!\!\!\! \sum_{(w,c)\in D\cup D'}\!\!\!\!\!\!\!\!\log \Big( \big(\frac{1}{1
+\exp^{-\textbf{v}_c^T\textbf{v}_w}}\big)^z\big(\frac{1}{1
+\exp^{\textbf{v}_c^T\textbf{v}_w}}\big)^{1-z}\Big)
\end{equation}
where $\textbf{v}_w\in \mathbb{R}^{n\times 1}$ and $\textbf{v}_c\in \mathbb{R}^{n\times 1}$
are the vectors of a word and its context respectively. $z$ indicates if this sample $(w,c)$ is a valid sample ($z=1$) or not ($z=0$)~\cite{Goldberg:2014}.
\cite{Pennington:2014} introduced global word vector (GloVe) as follows:
\begin{equation}
\underset{\theta} {\mathrm{arg~min}} \sum_{i,j}^{V} f(X_{ij}) (v_{w_i}^Tv_{w_j}+b_i+b_j-\log X_{ij}))    
\end{equation}

where $X_{ij} \in \mathbb{R}^{V\times V}$ is the co-occurrence matrix, $b_i$ and $b_j$ are constant biases, and $f(X)$ is defined as 
\begin{equation}
f(x) = \begin{cases}
             (x/x_{\text{max}})^{\alpha}  & \text{if } x < x_{\text{max}} \\
             1  & \text{other}
       \end{cases} 
\end{equation}
Estimating low-dimension vectors of sentences and documents is another interesting area in the litterateur. 
There are a couple of approaches to this end. 
The first approach is based on offline algorithms trying to estimate a document/sentence representation based on a number of available low-dimensional word vectors~\cite{Vulic:2015b,Clinchant:2013}. 
Usually a convolutional layer is used for the final estimation~\cite{Collobert:2011}.
The second approach is based on learning a representation vector for each document/sentence during learning the vectors of the words~\cite{V:2014}. 
Indeed, the proposed paragraph-to-vector is suitable for pre-defined documents or queries. 
Herein, we do not provide this approach first for lower performance of direct query-document low-dimensional vector similarity in ad-hoc retrieval and second for the streaming essence of the queries in this area.

Kusner et al., in \cite{Kusner:2015} exploited distributional document representations in a number of text classification tasks.
ALMasri et al., in \cite{ALMasri:2016} introduced a term-by-term expansion approach based on the low-dimensional vector similarities.
Kiros et al., in \cite{Kiros:2014} investigated on expanding a text by incorporating vector similarity of the candidates with averaged vector of the embedded words in the text.
The obtained vectors are used in sentiment classification, cross-lingual document classification, blog authorship attribution, and conditional word similarity. 
However, there is a number of works with advanced neural networks~\cite{Socher:2011,Socher:2013}. 
Socher et al., introduce recurrent neural networks (RNN) for sentence-formed queries/tweets.

\cite{Vulic:2015b} proposed an information retrieval framework based on the low-dimensional vectors. 
The authors demonstrated that the low-dimensional vectors are not yet effective in the vector-space ad-hoc retrieval frameworks where we compute document-query low-dimensional vector similarity.
Zheng et al., incorporated the word vectors in a supervised technique for re-weighting terms in probabilistic language model and BM25 \cite{Zheng:2015}.
Grbovic et al.,  uses query logs for building query models. The obtained models are used for query prediction and advertisement \cite{Grbovic:2015}.
A number of works investigate on using the embedded vectors in cross-lingual environments \cite{Dadashkarimi:2016,Vulic:2015b,Bengio:2014}.
\cite{Dadashkarimi:2016} employed an offline projection algorithm to bridge the gap between the languages. 
The authors incorporated the vector similarities for building a query language model.
\cite{Vulic:2015b} uses an on-line shuffling approach to this aim. 
The low-dimensional vectors are learned on a large comparable corpora after shuffling the words of each alignment. 
Dadashkarimi et al., demonstrated that the language model obtained by the projected vectors from different languages outperforms the shuffling approach \cite{Dadashkarimi:2016}.
Bengio et al., introduced BIBOWA, a fast on-line technique for learning multilingual word vectors, that learns the vectors from parallel corpora instead of bilingual dictionaries \cite{Bengio:2014}.

\section{Embedded Coefficients for Query Projection}
\label{Embedded Coefficients for Query Model Updating}
\begin{figure*}[t]
        \centering
        \includegraphics[width=0.65\textwidth]{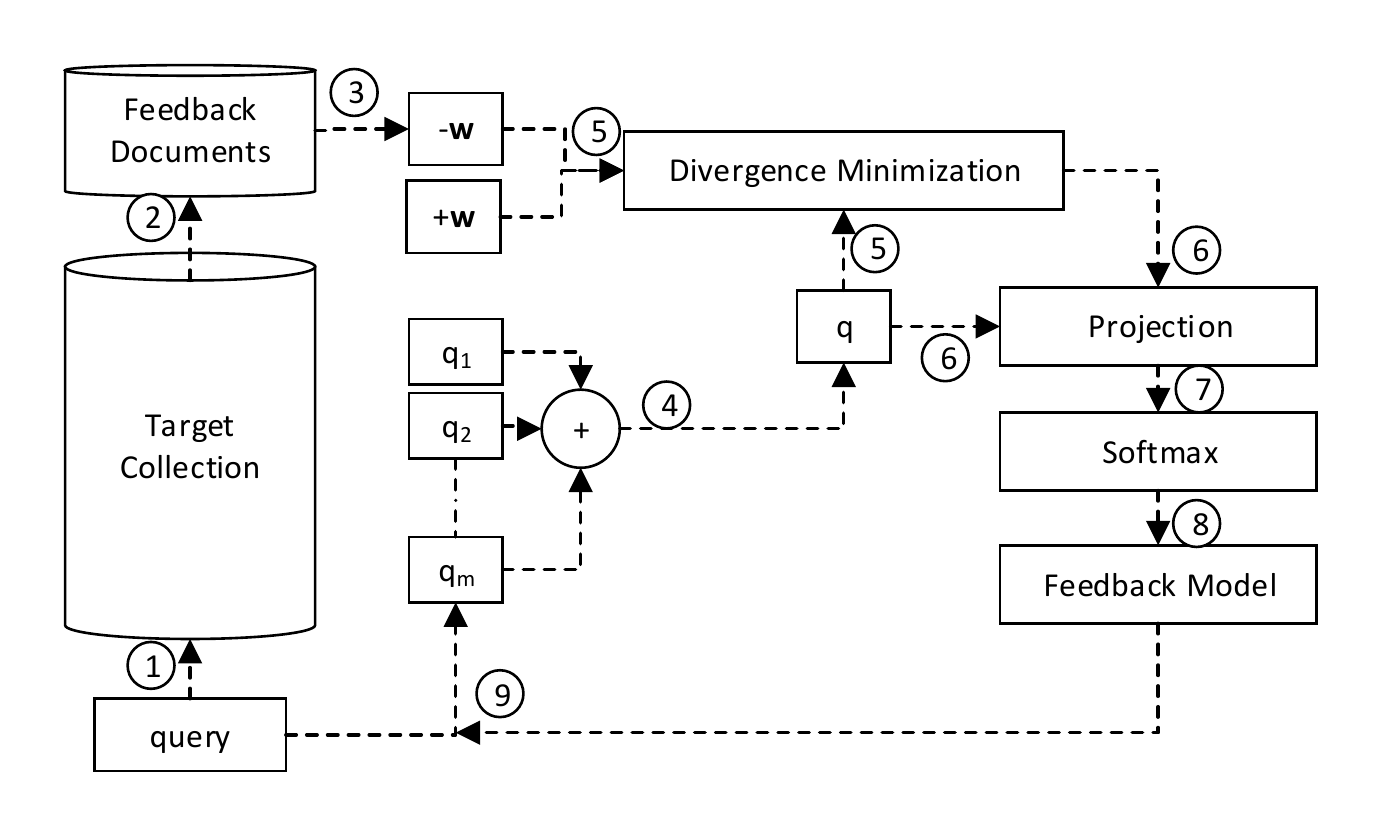}
        \caption{Whole process of low-dimensional query language modeling and feedback modeling.}
        \label{fig:whole}

\end{figure*}



In this section we propose a novel technique for query language modeling based on low-dimensional word vectors.
The proposed ECDMM takes the low-dimensional query vector and a number of relevant/irrelevant embedded vectors for its rotation and scaling. 
To this end, we aim to find a coefficient matrix $\textbf{W} \in \mathbb{R}^{n\times n}$ for projecting the query model $\textbf{v}_q$ to a more relevant space by minimizing $f$ as follows:
\begin{align}
\label{EQ:WEMEDMM}
    f(\textbf{W})=&\sum_{w_n\in F}\frac{\alpha}{2}||\textbf{W}^T\textbf{v}_q-\textbf{v}_{w_n}||^2 \nonumber\\
    -&\sum_{\bar{w}_n\in F}\frac{\lambda}{2}||\textbf{W}^T\textbf{v}_q-\textbf{v}_{\bar{w}_n}||^2 \nonumber\\
    -&\frac{\beta}{2}||\textbf{W}^T\textbf{W}||
\end{align}
where $\textbf{v}_q \in \mathbb{R}^{n\times 1}$ is the query vector, $\textbf{v}_{w_n}\in \mathbb{R}^{n\times 1}$ is the vector of a relevant sample from $F$, and $\textbf{v}_{\bar{w}_n}\in \mathbb{R}^{n\times 1}$ is the vector of a non-relevant sample from $F$.
$\alpha$ is a controlling constant for the positive samples and $\lambda$ is for the negative ones.
$\beta$ is another constant parameter for the regularization term $||\textbf{W}^T\textbf{W}||$.

The query vector is built from averaging the vectors of query words as follows:
\begin{equation}
[\textbf{v}_q]_j\leftarrow \frac{1}{m}\sum_{1\leq i \leq m=|q|}[\textbf{v}_{q_i}]_j     
\end{equation}
Equation~\ref{EQ:WEMEDMM} has similar components to MEDMM~\cite{Lv:2014}.
The first part captures the same essence as the cross-entropy between the feedback model $\theta_F$ and the positive sample model $\theta_w$ in MEDMM.
It tries to minimize the distance between the query model and the 
positive samples.
The second part also captures the effect of the negative samples on $\theta_F$. It tries to maximize the distance between the query model and the negative samples.
And, $\textbf{W}^T\textbf{W}$ acts as a regularization term for $\textbf{W}$ in the model.
Positive samples are drawn according to the following distribution:  
\begin{equation}
w^+\sim\frac{(1-\lambda)p_{\text{ml}}(w|\theta_F)}{(1-\lambda)p_{\text{ml}}(w|\theta_F)+\lambda p(w|\mathcal{C})}     
\end{equation}
where $\theta_F$ and $\theta_{\mathcal{C}}$ are unigram feedback distribution and unigram collection distribution respectively. $\lambda$ is set near to $0.9$ empirically for penalizing common words in $\mathcal{C}$ \cite{Zhai:2001}.
Negative samples are drawn as follows:
\begin{equation}
 w^-\sim p(w|\theta_\mathcal{C};F)^{\frac{3}{4}}    
\end{equation}
which is the unigram language model of the feedback documents raised to the power of $\frac{3}{4}$. 
This power is used for increasing the chance of appearing rare words  \cite{Goldberg:2014}. 
To find the optimum value of $\textbf{W}$ we use the stochastic gradient descent algorithm as follows:

\begin{equation}
     \textbf{W}^{t+1} = \textbf{W}^t - \eta \frac{\partial f}{\partial \textbf{W}}
\end{equation}

where $\eta$ is a constant learning rate.
Since $f(\textbf{W})$ is a differentiable single-variable quadratic function 
$\frac{\partial f}{\partial \textbf{W}}$ can be obtained as follows:
\begin{align}
\label{Eq:gradient descent}
     \frac{\partial f}{\partial \textbf{W}}&= \alpha\sum_{w_n\in F}(\textbf{W}^T\textbf{v}_q-\textbf{v}_{w_n})\textbf{v}_q^T \nonumber\\ &-\lambda\sum_{\bar{w}_n\in F} (\textbf{W}^T\textbf{v}_q-\textbf{v}_{\bar{w}_n})\textbf{v}_q^T \nonumber \\
     &-\beta \textbf{W}
\end{align}

Finally, the query vector $\textbf{V}_q$ can be projected to the new space as follows:
\begin{equation}
     \hat{\textbf{v}}_q = \textbf{W}^T\textbf{v}_q
\end{equation}
\begin{table*}[t]
\centering
\caption{Collections Characteristics}
\begin{tabular}{|c|c|c|c|c|c|c|} \hline
ID & Lang. & Collection & Queries (title only) & \#docs  & \#qrels \\ \hline
AP & English & Associated Press 88-89  & TREC 1-3 Ad-Hoc Track, Q:51-200 & 164,597 & 15,838\\ \hline
ROB356 & English &  \specialcell{Los Angeles Times 1994,  \\plus Glasgow Herald 1995} & \specialcell{CLEF 2003-2004-2006, \\Q:141-350} &  170,153& 4,327\\ \hline
SP & Spanish & EFE 1994 & CLEF 2002, Q:91-140 & 215,738 &  2,854\\ \hline
DE & German & \specialcell{Frankfurter Rundschau 94, \\SDA 94, Der Spiegel 94-95} & CLEF 2002-03, Q:91-140 & 225,371 &  1,938\\ \hline
FR & French & \specialcell{Le Monde 94, \\SDA French 94-95} & CLEF 2002-03, Q:251-350 & 129,806 & 3,524 \\ \hline
FA & Persian & Hamshahri 1996-2002 & CLEF 2008-09, Q:551-650 & 166,774 & 9,625 \\ \hline
\end{tabular}
\label{tab:dataset}
\end{table*}
\subsection{Building Embedded Query Model}
\label{Building Embedded Query Model}
In this subsection we shed light on a number of approaches for building a query language model.
Experiments in \cite{Vulic:2015b} demonstrate that using direct query-document vector similarity degrades the retrieval performance in monolingual document retrieval. 
Therefor, we aim to use a state-of-the-art retrieval framework as a core of document scoring and the low-dimensional vectors for query language modeling.

We believe that $\hat{\textbf{v}}_q$ represents a better semantic direction of the query and therefore, we aim to find a feedback model based on similarity of the projected query with the words from the feedback documents.
Herein, we examine a couple of functions to compute this similarity.
First we use the sigmoid function as follows:
\begin{align}
   \nabla_1(\hat{\textbf{v}}_q,\textbf{v}_{w_n}) = \frac{1}{1+e^{-\hat{\textbf{v}}_q\textbf{v}_{w_n}}}
   \label{eq:sigm}
\end{align}
Second we can use the cosine similarity to this end:
\begin{align}
   \nabla_2(\hat{\textbf{v}}_q,\textbf{v}_{w_n}) = \frac{\hat{\textbf{v}}_q\textbf{v}_{w_n}}{\|\hat{\textbf{v}}_q\|\|\textbf{v}_{w_n}\|}
   \label{eq:cosin}
\end{align}
To build a feedback language model we can use both $\nabla_1$ and $\nabla_2$ within a softmax layer as follows:
\begin{align}
    p(w_n|\hat{\theta}_F) = \frac{e^{\nabla(\hat{\textbf{v}}_q,\textbf{v}_{w_n})}}{\sum\limits_{w_n}e^{\nabla(\hat{\textbf{v}}_q,\textbf{v}_{w_n})}}
    \label{eq:IDF}
\end{align}

Equation \ref{eq:IDF} captures only specificity of the feedback terms \cite{Sparck:1972}. Term frequency is another useful metric for scoring the terms \cite{Sparck:1972}.
Herein, we use the following weighted softmax to incorporate both of these metrics:
\begin{align}
p(w_n|\hat{\theta}_F) \propto \frac{a_{w_n}e^{\nabla(\hat{\textbf{v}}_q,\textbf{v}_{w_n})}}{\sum\limits_{w_n}a_{w_n}e^{\nabla(\hat{\textbf{v}}_q,\textbf{v}_{w_n})}}
\label{eq:WIDF}
\end{align}
where $a_{w_n} = c(w_n,F)$ is the frequency of $w$ in $F$. 
The obtained feedback model can be interpolated with the original query as shown in Eq. \ref{Eq:linear}.

Figure \ref{fig:whole} shows whole the process of query language modeling and feedback modeling.




\section{Experiments}
\subsection{Experimental Setup}

\begin{table*}[t]
\caption{Investigating ECDMM performance using different vector similarity methods. $^*$ indicates the weighted softmax function introduced in Eq. \ref{eq:WIDF} (see also Eq. \ref{eq:sigm} and Eq. \ref{eq:cosin} for more details).}
\centering
\begin{tabular}{c|c|c|c||c|c|c|}
\cline{2-7}
        &   \multicolumn{3}{c||}{AP}  & \multicolumn{3}{c|}{ROB356} \\ \cline{2-7} \hline
      \multicolumn{1}{|c|}{ID}  & MAP    & P@5    & P@10    & MAP     & P@5     & P@10  \\ \hline
      \multicolumn{1}{|c|}{SIGMOID/$\nabla_1$}   &0.3136  &	0.4698 &0.4470 & 0.3434 & 0.4039 &	0.3667  \\ \hline
\multicolumn{1}{|c|}{COSIN/$\nabla_2$}   & 0.3150&	0.4711 & 0.4483 & 0.3486& 0.4183&0.3699	 \\ \hline
\multicolumn{1}{|c|}{SIGMOID$^*$/$\nabla_1$}   & 0.3241 & 0.4711 & 0.4544     & 0.3474    &  0.4078  & 0.3706    \\ \hline
\multicolumn{1}{|c|}{COSIN$^*$/$\nabla_2$}   &     \textbf{0.3330}   & \textbf{0.4792}      &   \textbf{0.4631}    &    \textbf{0.3866}     &  \textbf{0.4405}    &    \textbf{0.3810}    \\ \hline
\end{tabular}
\label{tab:delta}
\end{table*}

\begin{table*}[!htbp]
\caption{Comparison of different feedback methods. Superscripts 1/2/3/4/5 indicate that the MAP improvements are statistically significant compared to MLE/MIXTURE/RM3/RM4/MEDMM respectively (2-tail t-test, $p \leq 0.05$). The bold values in each column show the highest performance in terms of the corresponding metric for each collection.}
\centering
\begin{tabular}{c|c|c|c||c|c|c||c|c|c|}
\cline{2-10}
        &   \multicolumn{3}{c||}{AP}  & \multicolumn{3}{c||}{ROB356} & \multicolumn{3}{c|}{DE} \\ \cline{2-10} \hline
      \multicolumn{1}{|c|}{ID}  & MAP    & P@5    & P@10    & MAP     & P@5     & P@10   & MAP    & P@5   & P@10  \\ \hline
\multicolumn{1}{|c|}{MLE}  & 0.2643 & 0.451  & 0.4262    & 0.3721  & 0.4366  & 0.3719 & 0.348  & 0.532 & 0.458 \\ \hline
\multicolumn{1}{|c|}{MIXTURE} & 0.3106&	0.4450	& 0.4232 & 0.3781  & 0.4366  & 0.3758 & 0.4334	& 0.536& 0.488 \\ \hline
\multicolumn{1}{|c|}{RM3}     & 0.3187&	0.4470 & 0.4294 & 0.4037 &	0.4405&	0.3902 & 0.4346&	\textbf{0.5560}&	0.494 \\ \hline
\multicolumn{1}{|c|}{RM4}     & 0.2875 &	0.4208 &	0.3876 & 0.3789&	0.4392&	0.3752 & 0.3632&	0.5400 &	0.4720  \\ \hline
\multicolumn{1}{|c|}{MEDMM}   & 0.3269 &	0.4551	& 0.4289 & 0.3908&	\textbf{0.4523}&	0.3876 & 0.3878&	0.5400&	\textbf{0.4980} \\ \hline
\multicolumn{1}{|c|}{RFMF}   & 0.3296 &	0.4577& 0.4356& 	\textbf{0.4096} &	0.451 &\textbf{0.402}& 0.4248 &0.524	&0.488 \\ \hline
\multicolumn{1}{|c|}{ECDMM}   &     \textbf{0.3330}$^{12345}$   & \textbf{0.4792}      &   \textbf{0.4631}    &    0.3866$^{124}$     &  0.4405    &    0.3810   &    \textbf{0.4369}$^{145}$    &   0.552    &  0.4960  \\ 
\cline{2-10}\hline
        &   \multicolumn{3}{c||}{FA}  & \multicolumn{3}{c||}{FR} & \multicolumn{3}{c|}{SP} \\ \cline{2-10}\hline
      \multicolumn{1}{|c|}{ID}  & MAP    & P@5    & P@10    & MAP     & P@5     & P@10   & MAP    & P@5   & P@10  \\ \hline
\multicolumn{1}{|c|}{MLE}  & 0.3554 & 0.584 & 0.561 & 0.3936 & 0.5212 & 0.4556 & 0.488  & 0.664 & 0.578 \\ \hline
\multicolumn{1}{|c|}{MIXTURE} & 0.3934&	0.606&	0.577 & 0.4119&	0.499&	0.4616 & 0.5161&	0.644&	0.598 \\ \hline
\multicolumn{1}{|c|}{RM3}     & \textbf{0.4036}&	\textbf{0.618}&	\textbf{0.594} & 0.4115&	0.4889&	0.4556 & 0.5388&	0.6520&	0.6120 \\ \hline
\multicolumn{1}{|c|}{RM4}     & 0.3721 & 0.596 & 0.586 & 0.4031&	0.5172&	0.4606 & 0.5053&	0.6800&	0.594  \\ \hline
\multicolumn{1}{|c|}{MEDMM}   & 0.3585 & 0.59  & 0.56  & 0.4125&	\textbf{0.5253}&	0.4707 & 0.5268&	\textbf{0.688}&	0.608 \\ \hline
\multicolumn{1}{|c|}{RFMF}   & 0.392&	0.6100&\textbf{0.5940} & \textbf{0.4219} &	0.5051&0.4626	& \textbf{0.5459}& 0.6640&\textbf{0.6180}	 \\ \hline
\multicolumn{1}{|c|}{ECDMM}   &    0.3950$^{145}$ &  0.6040     & 0.5730  &  0.4217$^{12345}$    &     0.5152   &  \textbf{0.4717}      &   0.5384$^{1245}$     &  0.652     & 0.602\\ \hline
\end{tabular}
\label{tab:res_table}
\end{table*}

\begin{table}[!htbp]
	\centering
	\caption{Comparing ECDMM with word2vec and ECDMM with GloVe. The $*$ superscript shows a significant difference (2-tail t-test, $p \leq 0.05$).}
	\label{tab:glove}
	\begin{tabular}{c|c|c|c|c|}
		\cline{2-5}
		&\multicolumn{2}{|c|}{AP}     & \multicolumn{2}{|c|}{ROB356} \\
		\cline{2-5}
		&word2vec&GloVe &word2vec& GloVe
		\\ \hline
		\multicolumn{1}{|l|}{MAP} &\textbf{0.3330}$^*$&0.3245&0.3866& \textbf{0.3895}\\ \hline
		\multicolumn{1}{|l|}{P@5} &\textbf{0.4792}& 0.4738&\textbf{0.4405}& \textbf{0.4405}\\ \hline
		\multicolumn{1}{|l|}{P@10} &\textbf{0.4631}&0.4510&0.3810& \textbf{0.3856}\\ \hline
		
	\end{tabular}
\end{table}

\begin{figure*}[t]
	\centering
	\begin{subfigure}[b]{0.32\textwidth}
		\includegraphics[width=\textwidth]{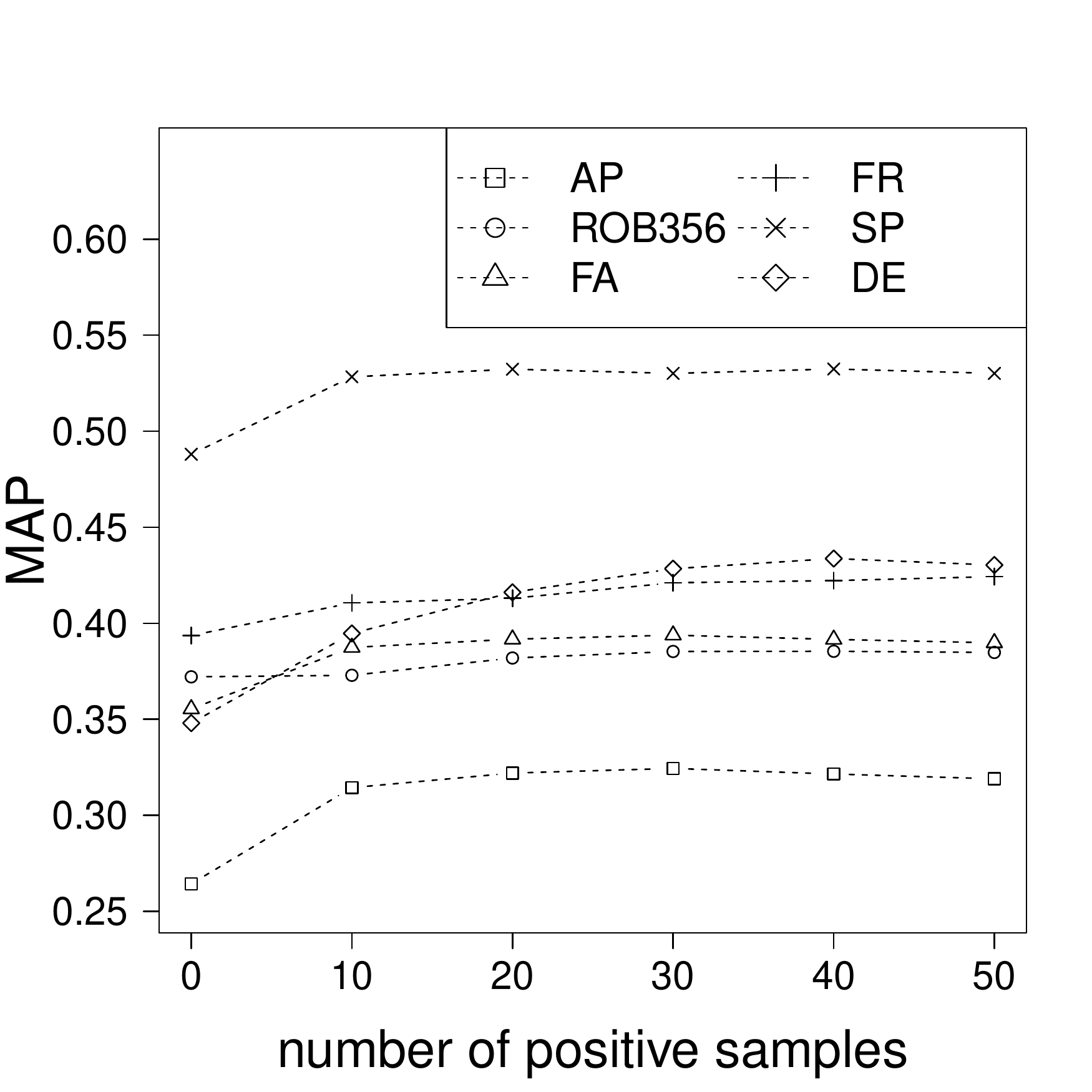}
	\end{subfigure}%
	~
	\begin{subfigure}[b]{0.32\textwidth}
		\includegraphics[width=\textwidth]{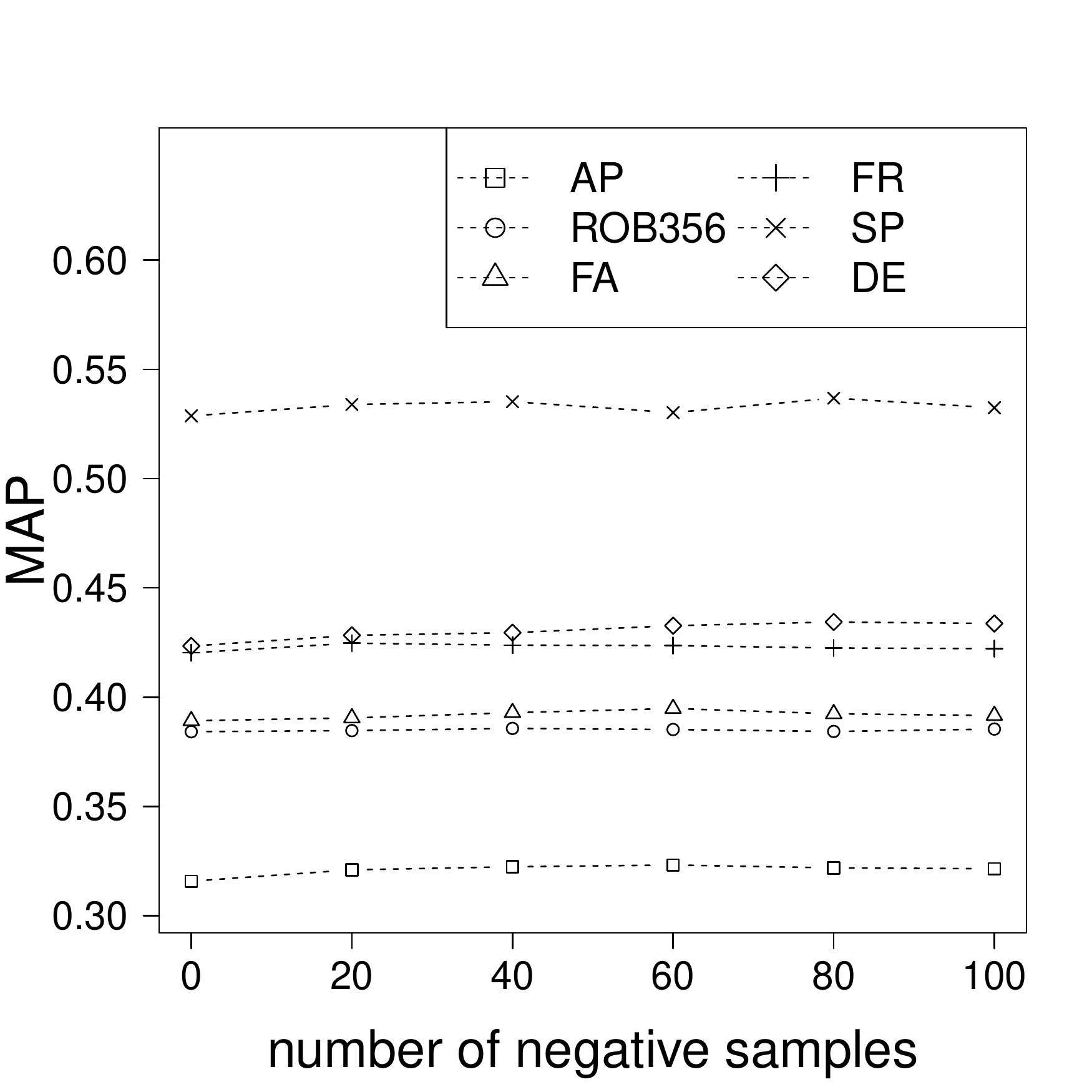}
	\end{subfigure}
	~ 
	\begin{subfigure}[b]{0.32\textwidth}
		\includegraphics[width=\textwidth]{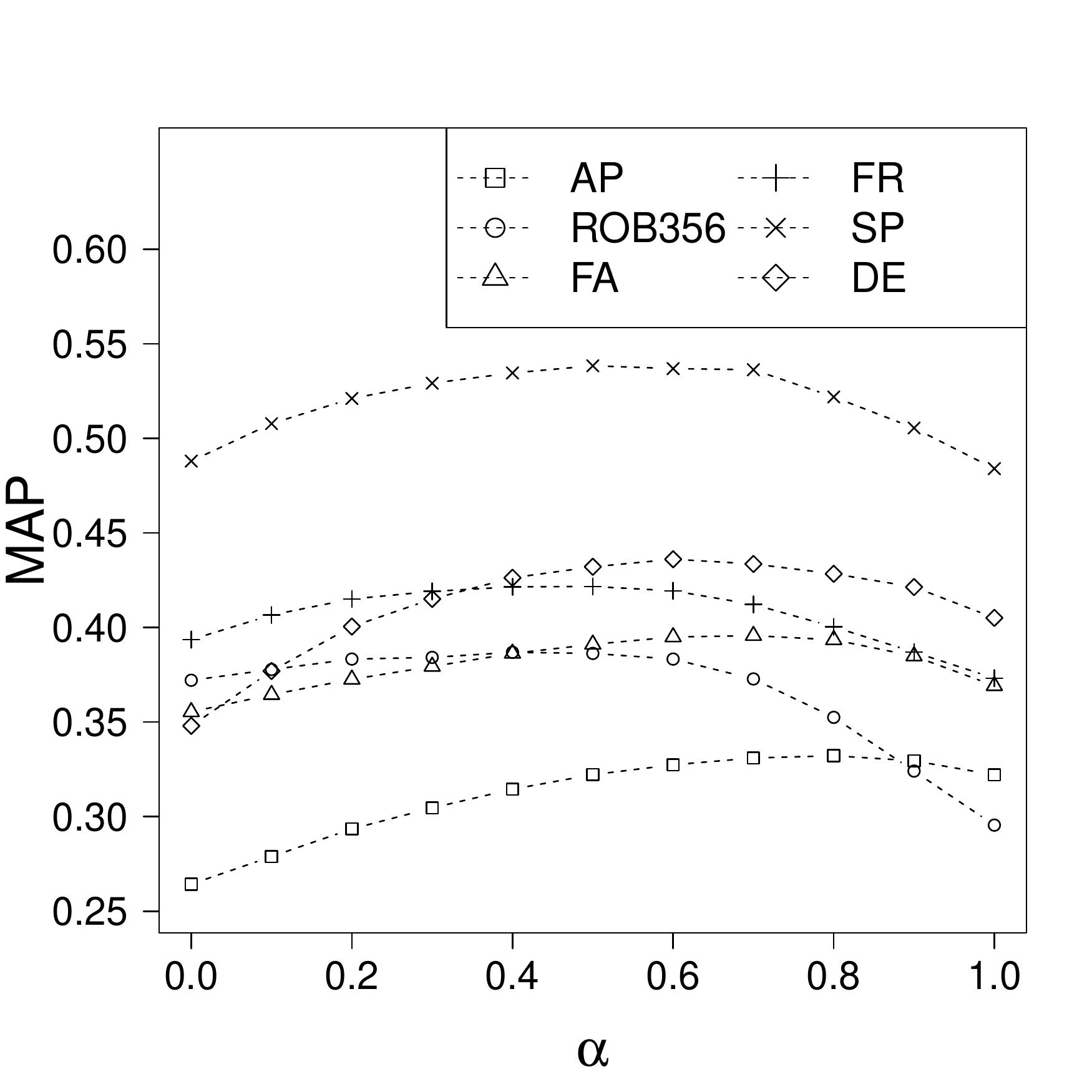}
	\end{subfigure}
	~
	\caption{MAP sensitivity of ECDMM to $n^+$, $n^-$, and $\alpha$ (feedback coefficient).}
	\label{fig:param}
\end{figure*}
\begin{table*}[t]
	\centering
	\caption{Query language models for \textit{"airbus subsidy"} by different feedback models. Terms are stemmed by the Porter stemmer.}
	\label{my-label}
	\begin{tabular}{|l|l||l|l||l|l||l|l||l|l||l|l|}\hline
		\multicolumn{2}{|c||}{MIXTURE}  & \multicolumn{2}{c||}{RM3}  & \multicolumn{2}{c||}{RM4}  & \multicolumn{2}{c||}{MEDMM} & \multicolumn{2}{c|}{RFMF}& \multicolumn{2}{c|}{ECDMM} \\ \hline
		airbu    & 0.0558 & airbu     & 0.0317 & said     & 0.0297  & airbu      & 0.1121 & airbu & 0.0299 & airbu      & 0.1001  \\
		subsidi  & 0.0274 & said        & 0.0197 & airbu       & 0.0246  & subsidi    & 0.0668  & subsidi& 0.0280 & subsidi    & 0.0610 \\
		us       & 0.0227  & airbu    & 0.0169  & subsidi    & 0.0179  & said& 0.0381 &consortium& 0.0226& industri   & 0.0203 \\
		govern   & 0.0219 & govern       & 0.0155 & s  & 0.01769 & us       & 0.0367 & manufactur& 0.0199 & aircraft   & 0.0173 \\
		trade     & 0.0167 & us & 0.0153 & us       & 0.01397  & govern     & 0.0326 & aircraft& 0.0170 & econom     & 0.0169  \\
		aircraft    & 0.0160  & s   & 0.0117 & govern   & 0.0137  & consortium & 0.0322 & germani& 0.0170& european   & 0.0158 \\
		west & 0.0158 & west     & 0.0110 & aircraft    & 0.0096 & aircraft      & 0.0271 & blohm& 0.0176 & yeutter    & 0.0140 \\
		industri  & 0.0156 & trade    & 0.0106 & trade    & 0.0093 & trade   & 0.0255 & boelkow& 0.0176& germani    & 0.0138 \\
		european  & 0.0132 & industri     & 0.0095  & will     & 0.0059 & west & 0.0088  & spain& 0.0160 & daimler    & 0.0137 \\
		consortium   & 0.0126 & aircraft    & 0.0094 & aircraft & 0.0052 & state       & 0.0085 & yeutter & 0.0160& consortium & 0.0126 \\ \hline
	\end{tabular}
	\label{tab:expansion terms}
\end{table*}


Overview of the used collections are provided in Table~\ref{tab:dataset}.
In all experiments, we use the language modeling framework with the KL-divergence retrieval model and Dirichlet smoothing with $\mu=1000$.
All European documents and queries are stemmed by the Porter stemmer and the Persian collection is remained intact~\cite{Hashemi:2014,Rahimi:2016}.
Therefore, we might have different vectors for cognate words.
Stopwords are removed in all the experiments\footnote{We use the stopword lists and the normalizing techniques available at \url{http://www.unine.ch/info/clef/}.}.
The Lemur toolkit\footnote{\url{http://www.lemurproject.org/}} is employed as the retrieval engine in our experiments.
ECDMM is compared with the following methods:  (\textit{1}) maximum likelihood estimation of query (MLE): $p(w|\theta_q)\!\!=\!\!\frac{c(w,q)}{|q|}$ where $c(w,q)$ is the count of term $w$ in the query; (\textit{2}) RM3, (\textit{3}) RM4, and (\textit{4}) MEDMM explained in Section~\ref{Sec:Pseudo-relevance Feedback}. 

$\alpha$ in Equation~\ref{Eq:linear} is set via 2-fold cross validation over topics of each collection and number of blind relevant documents is assumed $|F|\!\!=\!\!10$.
All free parameters $\alpha,\lambda, \beta, n^+, \text{and } n^-$ are fixed for all experiments after learning on a small sub-set of topics from the AP collection.
Empirically we fixed the parameters to $\alpha\!\!=\!\!0.8$, $\lambda\!\!=\!\!0.05$, $\beta\!\!=\!\!0.01$, $n^+\!\!=\!\!40$, and $n^-\!\!=\!\!100$ in all the experiments.
$\textbf{W}$ in Equation~\ref{Eq:gradient descent} is initialized with random values in $[-1,1]$;
$\eta$ is set to a small value which also decreases after each iteration.
The iterations terminate when the changes are very small or the number of iterations meets $1000$.

The words' vectors computed with word2vec introduced in~\cite{Mikolov:2013}; size of the window, number of negative samples, and size of the vectors are set to typical values of $10$, $45$, and $100$ respectively.
Vectors of GloVe are extracted from Wikipedia and Gigawords 5 with 6 billion tokens and 400k words \footnote{\url{http://nlp.stanford.edu/projects/glove/}}.
The number of dimensions in GloVe is set $100$ similarly.
\subsection{Comparing word2vec and GloVe}

In this section we aim to choose a standard low-dimensioanl vectors for our primary experiments.
Therefore, we investigate performance of the proposed method using both word2vec and GloVe \cite{Mikolov:2013,Pennington:2014}.
Query vectors are built using the vectors obtained from word2vec and GloVe for the query terms in AP and ROB356.
After that, the same configuration is used for evaluating the performance of ECDMM based on these vectors.

As shown in Table \ref{tab:glove} ECDMM with word2vec outperforms ECDMM with GloVe in AP significantly.
Differences in ROB356 are not statistically significant and thus we opted word2vec vectors in the rest of experiments.

\subsection{Performance Comparison and Discussion}

Table \ref{tab:delta} shows experimental results on different vector similarity methods. 
According to the results, cosine similarity with a weighted softmax function works better than other similarity metrics.
Indeed, the experimental results demonstrate that considering only the vector similarity captures the IDF value of the terms.
But, incorporating this with counts of the terms works more better.
Therefor we opt cosine similarity with the weighted softmax function in the rest of experiments for comparison and parameter sensitivity.

All the experimental results are provided in Table~\ref{tab:res_table}. 
As shown in the table, the proposed ECDMM outperforms a few number of baselines in terms of MAP and works as well as the state-of-the-art feedback language models.
The results show 26.0\%, 3.8\%, 25.5\%,  11.1\%, 7.1\%, 10.3\% improvements in AP, ROB356, DE, FA, FR, and SP respectively in terms of MAP compared to MLE.
In AP and DE, ECDMM outperforms all the baselines in terms of all the metrics.
Generally, the results in P@5 and P@10 are very competitive and there is not any specific method that outperforms all other ones in all the collections.

RFMF outperforms all the baselines and the proposed ECDMM in SP in all metrics. Differences in this collection are statistically significant. 
On the other hand, ECDMM outperforms RFMF in AP and DE significantly.
Differences in other collections are not statistically significant.

As stated before, we have not used any stemmer for Persian and thus we might have different vectors for cognate words.
But, the results belonging to FA demonstrate that the obtained vectors are suitable enough for our model.

As discussed in Section~\ref{Embedded Coefficients for Query Model Updating}, ECDMM takes advantage of the first step of MIXTURE for positive sampling and the idea of MEDMM for divergence minimization. 
The experimental results reveal that ECDMM is more effective than these methods and captures both topicality and entropy.
However, RM3 is a strong baseline and works better than the proposed method in ROB356, FA, and SP in terms of MAP.
It is noteworthy that $p(\theta_d)$ in RM1/RM3 (see Eq. \ref{eq:rm1}) plays key role in its efficiency and it would be interesting to study this effect on $a_{w_n}$ in Eq. \ref{eq:WIDF}.
Nevertheless, the differences are not statistically significant.

Table~\ref{tab:expansion terms} shows top-10 stemmed expansion terms and the weights of obtained query model by the methods for \textit{'airbus subsidy'} from topics of AP.
It shows that MIXTURE and ECDMM are more successful in purifying the feedback model from common words (see \textit{s} and \textit{said} in the lists).
However, terms like \textit{us} never appeared in ECDMM although it is semantically related to the query;
As shown in the table, MEDMM and ECDMM weight the original query more than others (see \textit{airbu} and \textit{subsidi} in the lists). 
Therefor, higher value of $\alpha$ works well for the proposed model in the AP collection (see Fig.\ref{fig:param} and \cite{Lv:2014}).

\subsection{Parameter Sensitivity}
In this section we investigate the sensitivity of the proposed method to the number of positive and negative samples and the feedback coefficient (see Eq. \ref{Eq:linear}).
To this aim, one parameter is fixed to its optimum value and different values are tested for the other one.
As shown in Figure~\ref{fig:param} both parameters $n_+$ and $n_-$ work stable in all the collections.
However as shown in the figure, the performance of the system is less reliable to $n_-$ than $n_+$.
The reason might be due to existing a variety of topics in the non-relevant low-dimensional space.
Nevertheless, in difficult queries, which is not the focus of the current work, it is necessary to consider negative feedback as well.
In this kind of circumstances, $F$ contains fewer number of relevant documents and a retrieval system is required to use this negative information for query modification~\cite{Wang:2008}.

Vulic et al., have shown that the retrieval performance is not sensitive to the number of dimensions considerably \cite{Vulic:2015b}.
Therefor, herein we fixed this parameter to the typical value of $100$ and investigate sensitivity of other parameters.

\section{Conclusion and Future Works}
In this paper, we propose a query language model using low-dimensional query projection.
We use a set of positive and negative samples from the top-ranked documents, retrieved by the query, to learn an embedded coefficient matrix.
The query vector, which got transformed by the coefficient matrix, is then used to expand the original query.
We tested a couple of cosine and sigmoid functions for computing vector similarity of the projected vector and the feedback terms. 
The experimental results reveal that using the cosine similarity and a softmax layer works as well as the state-of-the-art feedback techniques and even better in a few number of collections.
ECDMM has significant improvements up to 3.8\% compared to the state-of-the-art models in MAP.

This work inspires a number of future works. First, we want to study the usage of the proposed method in low-dimensional profile modelling in recommendation systems.
In recomender systems, there is stream of documents being proposed to the users and thus we can update the low-dimensional profile vectors incrementally.
The proposed formulation of low-dimensional query updating can be adapted to this work.
Although the main focus of this work is to provide a robust formulation for query modeling/updating, tailoring semantic networks (e.g., WordNet, Concept-Net.) for positive sampling seems to be interesting.
Therefor, our second goal is to study the effect of negative sampling on difficult queries.

\section{Acknowledgment}
The authors would like to thank Hamed Zamany from Google corporation for his helpful comments on this work.





\bibliography{emnlp2016}
\bibliographystyle{abbrv}

\end{document}